\documentclass[twocolumn, secnumarabic, amssymb, nobibnotes, aps,superscriptaddress]{revtex4-1} 
\usepackage{amsmath,amstext,bm,appendix,graphicx,multirow}
\usepackage{color}
\usepackage{diagbox}

\begin{document}
\title{Construction of robust Rydberg controlled-phase gates}
\author{C. P. Shen}
\affiliation{School of Physics and Engineering, Zhengzhou University, Zhengzhou 450001, China}
\author{J. L. Wu}
\affiliation{Department of Physics, Harbin Institute of Technology, Harbin 150001, China}
\author{S. L. Su}
\email[]{slsu@zzu.edu.cn}
\affiliation{School of Physics and Engineering, Zhengzhou University, Zhengzhou 450001, China}
\author{E. J. Liang}
\affiliation{School of Physics and Engineering, Zhengzhou University, Zhengzhou 450001, China}

\begin{abstract}
One scheme is presented to construct the robust multi-qubit arbitrary-phase controlled-phase gate (CPG) with one control and multiple target qubits in Rydberg atoms using the Lewis-Riesenfeld (LR) invariant method. The scheme is not limited by adiabatic condition while preserves the robustness against control parameter variations of adiabatic evolution. Comparing with the adiabatic case, our scheme does not require very strong Rydberg interaction strength. Taking the construction of two-qubit $\pi$ CPG as an example, our scheme is more robust against control parameter variations than non-adiabatic scheme and faster than adiabatic scheme.
\end{abstract}
\maketitle

Rydberg atoms, the neutral atoms with high-lying Rydberg states, possess the stable ground state hyperfine levels, long coherence time of Rydberg state, and strong Rydberg-Rydberg interaction (RRI), which are promising candidates to be used to construct quantum logic gates~\cite{t001}. If one atom is excited to a high-lying Rydberg state, RRI~\cite{t002,t0021} would shift the Rydberg states in the surrounding atoms, i.e., excitation of the surrounding atoms would be inhibited. This phenomenon is called Rydberg blockade mechanism which has been demonstrated~\cite{t003,t004} and utilized to realize quantum logic gates~\cite{t005} in experiment. To suppress the errors of control parameter fluctuations, adiabatic evolution~\cite{t006} technology is usually adopted to accomplish the quantum information processing tasks in Rydberg atoms~\cite{t007,t0071}. {Nevertheless,} the adiabatic evolution is limited by adiabatic condition which leads to a slow evolution speed and would be easily spoiled by decoherence noises. Recently, shortcut to adiabaticity (STA) is studied to accelerate the adiabatic evolution while preserves the advantage of robustness, such as transitionless quantum driving~\cite{t009,t010}, Lewis-Riesenfeld (LR) invariant theory~\cite{t011,t0110,t0120,t012}. In Ref.~\cite{t013}, transitionless quantum driving is discussed for preparation of mutiple-Rydberg-atom entangled state.

In this letter, we use the STA of LR invariant method to construct robust multi-qubit Rydberg arbitrary-phase controlled-phase gate (CPG) with one control and multiple target qubits which can be applied to quantum error correction~\cite{t016}, discrete cosine transform~\cite{t017}. Analyses show that the adiabatic scheme requires very strong RRI strength to construct CPG while our three-step STA scheme performs well with a smaller RRI strength. The STA scheme is not only faster than adiabatic scheme, but also more robust against control parameter variations than non-adiabatic scheme.

\begin{figure}
  \centering
  \includegraphics[scale=0.8]{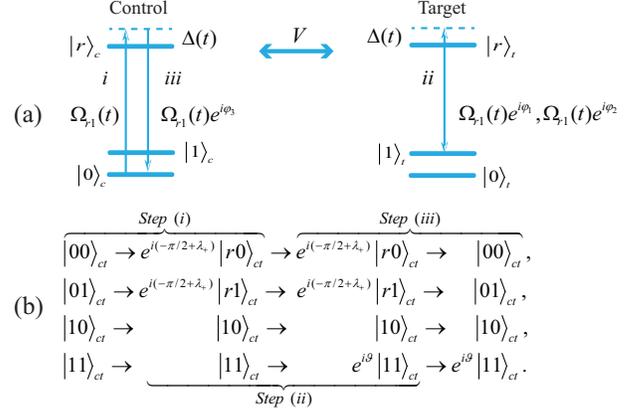}\\
  \caption{(a) Configuration for constructing two-qubit arbitrary-phase CPG.
  (b) State evolutions in the three-step scheme.}\label{fig1}
\end{figure}

Considering two Rydberg atoms with vdW interaction Hamiltonian (set $\hbar=1$) $H_{r} =V|rr\rangle\langle rr|$ in Fig.~\ref{fig1}(a), each atom consists of two stable ground states $|0\rangle$, $|1\rangle$ and one Rydberg state $|r\rangle$. $|0\rangle_{c}$ and $|1\rangle_{t}$ are coupled to $|r\rangle_{c}$ and $|r\rangle_{t}$ by Rabi frequencies $\Omega_{r1}(t)$ [$\Omega_{r1}(t)e^{i\varphi_{3}}$] and $\Omega_{r1}(t)e^{i\varphi_{1}},\Omega_{r1}(t)e^{i\varphi_{2}}$, respectively (here and hereinafter, subscript \emph{c} denotes control atom while \emph{t} denotes target atom), {where} $\varphi_{1},\varphi_{2},\varphi_{3}$ {are time-independent laser phases}. Assuming these two atoms have the same detuning $\Delta(t)=\omega-\omega_{0}$, where $\omega$ {denotes the} laser frequency and $\omega_{0}$ {denotes the} atomic transition frequency. In the interaction picture, Hamiltonian of the Rydberg atom reads
\begin{equation}\label{eq1}
  H=\frac{1}{2}[\Omega_{r1}(t)(|r\rangle\langle a|e^{-i\varphi}
  +\text{H.c})+\Delta(t)
  (|a\rangle\langle a|-|r\rangle\langle r|)],
\end{equation}
{where $|a\rangle$ is $|0\rangle_{c}$ in steps (i) and (iii) while is $|1\rangle_{t}$ in step (ii)}. The instantaneous eigenstates are $|\Phi_{+}(t)\rangle=\cos(\theta/2)|a\rangle+\sin(\theta/2)e^{-i\varphi}|r\rangle$, $|\Phi_{-}(t)\rangle=-\sin(\theta/2)e^{i\varphi}|a\rangle+\cos(\theta/2)|r\rangle$, where the mixing angle $\theta=\arccos[\Delta(t)/\Omega(t)]$ with $\Omega(t)=\sqrt{\Delta^2+\Omega_{r1}^2}$, and the corresponding eigenvalues {are} $E_{\pm}=\pm\Omega(t)/2$. For $\varphi=0$, {the} initial state $|\Phi_{+}(0)\rangle$ or $|\Phi_{-}(0)\rangle$ would evolve along the corresponding eigenstate, picking up {an} adiabatic phase $\varepsilon_{\pm}$ including~\cite{t014} dynamical component $-\int_{0}^{t}E_{\pm}(t')dt'$ and geometric component $i\int_{0}^{t}\langle\Phi_{\pm}(t)|\partial_{t'}\Phi_{\pm}(t)\rangle dt'$, when the adiabatic condition $|(\Omega_{r1}\dot{\Delta}-\dot{\Omega}_{r1}\Delta)/\Omega^3|\ll1$ is met (dot denotes time derivative).

{To realize the STA with LR invariant method in {a} two-level system, an invariant Hermitian operator}~\cite{t012}
\begin{equation}\label{eq2}
  I(t)=\frac{\chi}{2}[\cos\alpha(|a\rangle\langle a|-|r\rangle\langle r|)+\sin\alpha(|r\rangle\langle a|e^{-i\beta}+\text{H.c})]
\end{equation}
{is required which meets} $i\hbar\partial_{t}I(t) = [H(t), I(t)]$~\cite{t011}. Its eigenvalues $\eta_{\pm}=\pm\chi/2$ ($\chi$ is an arbitrary constant with units of frequency) and {instantaneous} eigenstates
\begin{eqnarray}\label{eq21}
  |\phi_{+}(t)\rangle &=& \cos(\alpha/2)|a\rangle+\sin(\alpha/2)e^{-i\beta}|r\rangle, \nonumber\\
  |\phi_{-}(t)\rangle &=& -\sin(\alpha/2)e^{i\beta}|a\rangle+\cos(\alpha/2)|r\rangle,
\end{eqnarray}
where $\alpha(t)$ and $\beta(t)$ are two time-dependent parameters.

{Since the initial states $|\phi_{\pm}(0)\rangle$ would evolve along the corresponding eigenstates $|\phi_{\pm}(t)\rangle$~\cite{t0120}, the solution of Schr\"{o}dinger equation can be written as $\Sigma_{+,-}C_{\pm}e^{i\lambda_{\pm}}|\phi_{\pm}(t)\rangle$, where $C_{\pm}$ are constant amplitudes, $\lambda_{\pm}=1/2\int_{0}^{t_{f}}[\Delta(t')-2\tilde{\Omega}(t')]dt'$ are LR phases with $\tilde{\Omega} =(\Delta+\dot{\beta})\cos^2\left(\alpha/2\right)
+(\Omega_{r1}/2)\sin\alpha\cos(\beta-\varphi)$}.
To meet the {invariant} definition equation $i\hbar\partial_{t}I(t) = [H(t), I(t)]$, according to Eqs.~(\ref{eq1}) and (\ref{eq2}), $\Omega_{r1}$ and $\Delta$ are derived as
\begin{eqnarray}\label{eq3}
  \Omega_{r1} &=& \dot{\alpha}/\sin(\beta-\varphi), \nonumber\\
  \Delta &=& \Omega_{r1}\cot\alpha\cos(\beta-\varphi)-\dot{\beta}.
\end{eqnarray}
{If we set the commutation relation $[H(0),I(0)]=[H(t_{f}),I(t_{f})]=0$~\cite{t012}, $H(t)$ and $I(t)$ would share common eigenstates at the begining and ending of {the systematic} evolution process, where $t_{f}$ {denotes the} interaction time. Thus the STA scheme based on LR invariant is not limited by adiabatic condition while can achieve the same purpose with adiabatic scheme.}

Next we focus on constructing two-qubit arbitrary-phase CPG $(|00\rangle\langle00|+|01\rangle\langle01|+|10\rangle\langle10|+e^{i\vartheta}|11\rangle\langle11|)$ with $\vartheta\in[-\pi,~\pi]$ {based on the STA of LR invariant method}. The construction process {requires} three steps. {Firstly, controlling the Rabi frequency $\Omega_{r1}(t)$ coupling $|0\rangle_{c}$ with $|r\rangle_{c}$ to realize the evolution of \emph{Step (i)} in Fig.~\ref{fig1}(b)}. {Secondly, controlling Rabi frequencies $\Omega_{r1}(t)e^{i\varphi_{1}},\Omega_{r1}(t)e^{i\varphi_{2}}$ successively, coupling $|1\rangle_{t}$ with $|r\rangle_{t}$ to realize the evolution of \emph{Step (ii)} in Fig.~\ref{fig1}(b)}. {Thirdly, controlling Rabi frequency $\Omega_{r1}(t)e^{i\varphi_{3}}$ coupling $|0\rangle_{c}$ with $|r\rangle_{c}$ again to realize the evolution of \emph{Step (iii)} in Fig.~\ref{fig1}(b)}. We will describe these three steps in detail below.

\emph{Step (i)}. {When the initial control qubit is} $|0\rangle_{c}$, with $H(t)$ in Eq.~(\ref{eq1}), {it will be excited to $|r\rangle_{c}$ along one of the two invariant eigenstates [here we consider the eigenstate $|\phi_{+}(t)\rangle$ in Eq.~(\ref{eq21}) as an example to illustrate the process]}, up to one phase factor labeled $e^{-i\beta_{1}}$. The LR amplitudes $C_{+}=\langle\phi_{+}|0\rangle=1$, $C_{-}=0$ and LR phase will be considered later. To deduce $H(t)$, commutation relation between $H(t)$ and $I(t)$ {needs to be satisfied at the begining and ending of the evolution process}~\cite{t012}. Besides, if $\Omega_{r1}(0)=\Omega_{r1}(t_{f})=0$, it is beneficial for experimental realization. To meet the above conditions, {setting}
\begin{equation}\label{eq51}
  \alpha(0)=0,~\alpha(t_{f})=\pi,~\dot{\alpha}(0)=\dot{\alpha}(t_{f})=0.
\end{equation}
For simplicity, {setting} $\varphi=0$, it is useful to make $\beta$ close to $(n+1/2)\pi$ to minimize $\Omega_{r1}$ ($n=0,1,2...$) in Eq.~(\ref{eq3}). And it is better for positive $\Omega_{r1}$ and definite $\Delta$. Thus $\beta$ should satisfy~\cite{t012}
\begin{equation}\label{eq4}
  \beta(0)=\beta(t_{f})=\beta_{1}=\pi/2,~\dot{\beta}(0)=-\dot{\beta}(t_{f})=3\pi/2t_{f}
\end{equation}
to meet the above conditions.
With boundary conditions {in} Eqs.~(\ref{eq51}) and (\ref{eq4}), for polynomial ansatz of $\alpha(t)=\sum_{j=0}^{ 3}a_{j}t^{j}$, $\beta(t)=\sum_{j=0}^{3}b_{j}t^{j}$, $\alpha$ and $\beta$ can be easily obtained. Then the control parameters $\Omega_{r1}(t)$ and $\Delta(t)$ {of Hamiltonian $H(t)$ in step(i)} would be derived.

After step (i), Transformations $|00\rangle_{ct}\rightarrow e^{i(-\beta_{1}+\lambda_{+})}|r0\rangle_{ct}$ and $|01\rangle_{ct}\rightarrow e^{i(-\beta_{1}+\lambda_{+})}|r1\rangle_{ct}$ are achieved [as shown in Fig.~\ref{fig1}(b)] while $|10\rangle_{ct},|11\rangle_{ct}$ remain unchanged, where $\lambda_{+}$ is LR phase.

\emph{Step (ii)}. {Turning on the laser that interacts with the target atom with the Hamiltonian shown in Eq.~(\ref{eq1}), which may have different parameter conditions from $H(t)$ {used} in step (i)}. {In step~(ii)}, evolution $|1\rangle_{t}\rightarrow|1\rangle_{t}e^{i\vartheta}$ {would be realized provided that the state of the control atom is initially in $|1\rangle_{c}$}.
For clearness, the {whole} physical process of step~{(ii)} can be classified as two types:

Type one: {When the control and target atoms are initially in states $|1\rangle_{c}$ and $|1\rangle_{t}$, respectively}, $|1\rangle_{c}$ would not be excited after step~(i). {And} $|1\rangle_{t}$ {will} be excited to $|r\rangle_{t}$ along one of the invariant eigenstates [consider $|\phi_{+}(t)\rangle$ in Eq.~(\ref{eq21})] {induced by} $H(t)$, {accompanied by a} phase factor $e^{-i\beta_{2}}$, which is similar to step~(i). Note that Eqs.~(\ref{eq51}) and (\ref{eq4}) are still {required} {and here labeling} $\beta(0)=\beta(t_{f})=\beta_{2}$. {Besides, time-independent} laser phase $\varphi_{1}$ {should satisfy} $\beta_{2}-\varphi_{1}=\pi/2$ to {make the} Rabi frequency $\Omega_{r1}(t)$ and detuning $\Delta(t)$ {the same as that in step (i)}. Next, {letting} $H(t)$ {works} again {and labeling the time-independent} laser phase {as} $\varphi_{2}$, $|r\rangle_{t}$ would be driven to $-|1\rangle_{t}$ along the eigenstate $|\phi_{-}(t)\rangle$ ($C_{+}=0$, $C_{-}=1$), {accompanied by a} phase factor $e^{i\beta_{3}}$. Here, {setting} $\beta(0)=\beta(t_{f})=\beta_{3}$ and $\beta_{3}-\varphi_{2}=\pi/2$ {to obtain the same Rabi frequency $\Omega_{r1}(t)$ and detuning $\Delta(t)$ in step (i)}.

Type two: {When the control qubit is initially in} $|0\rangle_{c}$, it would be excited {to Rydberg state} after step~(i). Thus the population {transfer} $|1\rangle_{t}\rightarrow|r\rangle_{t}$ would be inhibited due to Rydberg blockade~\cite{t002,t015} {with the condition $V\gg\Omega_{r1}$}.

After step (ii), transformation $|11\rangle_{ct} \xrightarrow[]{\Omega_{r1}e^{i\varphi_{1}}} \mathcal{A}|1r\rangle_{ct} \xrightarrow[]{\Omega_{r1}e^{i\varphi_{2}}}\mathcal{A}\mathcal{B}|11\rangle_{ct}$ is achieved while $|r0\rangle_{ct},|r1\rangle_{ct},|10\rangle_{ct}$ remain unchanged~{[as shown in Fig.~\ref{fig1}(b)]}, where $\mathcal{A}=e^{i(-\beta_{2}+\lambda_{+})}$, $\mathcal{B} = e^{i(\pi+\beta_{3}+\lambda_{-})}$ and $\lambda_{\pm}$ are LR phases. Then arbitrary phase $\vartheta=\pi+\beta_{3}-\beta_{2}$ is obtained since $\lambda_{-}=-\lambda_{+}$.

\emph{Step (iii)}. {With the Hamiltonian $H(t)$ in step (i) and labeling the time-independent} laser phase $\varphi_{3}$, state $|r\rangle_{c}$ would be driven to $-|0\rangle_{c}$ along $|\phi_{-}(t)\rangle$, {accompanied by a} phase factor $e^{-i\beta_{1}}$. Here it requires $\varphi_{3}$ to be $\pi$ to obtain the same {Rabi frequency $\Omega_{r1}(t)$ and detuning $\Delta(t)$} in step (i), and more important{ly}, to offset the LR phase $\lambda_{+}$ in step (i) ($\lambda_{-}=-\lambda_{+}$). After step (iii), $|r0\rangle_{ct}\rightarrow-e^{i(-\beta_{1}+\lambda_{-})}|00\rangle_{ct}$ and $|r1\rangle_{ct}\rightarrow-e^{i(-\beta_{1}+\lambda_{-})}|01\rangle_{ct}$ are achieved {as shown} in Fig.~\ref{fig1}(b) while $|10\rangle_{ct},|11\rangle_{ct}$ remain unchanged.
Considering these three steps, two-qubit arbitrary-phase CPG $(|00\rangle\langle00|+|01\rangle\langle01|+|10\rangle\langle10|
+e^{i\vartheta}|11\rangle\langle11|)$ is realized with $\vartheta\in[-\pi,~\pi]$. The total operation time is $4t_{f}$.

The scheme can be generalized to construct multi-qubit arbitrary-phase CPG with one control {and} multiple target qubits {in three steps}. Suppose {the control atom   is the same as that in Fig.~\ref{fig1}(a)} and all {of the} target atoms {are the same as the target atom} in Fig.~\ref{fig1}(a). {Then}, the RRI Hamiltonian {is changed to} $H_{r}' = \sum_{j=2}^{n}V\sigma_{1}^{rr}\otimes\sigma_{j}^{rr}
+\sum_{j,j'}^{n}V_{1}\sigma_{j}^{rr}\otimes\sigma_{j'}^{rr}$, where $j~(j')=2,3,\ldots, n~(j<j')$, $\sigma_{k}^{mn}$ stands for $|m\rangle\langle n|$ of the \emph{k}-th atom ($k=1$ for control atom, while $j,~j'$ for target atoms). $V_{1}$ denotes {RRI} strength between target atoms. {Considering the full Hamiltonian with the condition $V\gg\Omega_{r1}\gg V_{1}$~\cite{t0181}}, \emph{n}-qubit arbitrary-phase CPG
\begin{equation}\label{eq6}
  |\mu\nu\cdots\zeta\rangle_{123\cdots n} \rightarrow e^{i\Theta}|\mu\nu\cdots\zeta\rangle_{123\cdots n}
\end{equation}
would be constructed {with operation steps the same as that of the two-qubit case},  where $\Theta =\mu(\nu\vartheta_{1}+\cdots+\zeta\vartheta_{n-1})$, and $\mu,\nu,\cdots,\zeta\in\{0,~1\}$ while $ \vartheta_{1},\vartheta_{2},\cdots,\vartheta_{n-1}\in[-\pi,~\pi]$.

{We take the two-qubit $\pi$ CPG (labeled $U_{1}$) as an example to show the robustness against control parameter variations [set laser phase $\varphi_{1}=\varphi_{2}=0$ in step (ii)], and discuss its performance comparing with adiabatic and non-adiabatic cases}.

\emph{Adiabatic case}. Similar to the STA scheme, it needs three steps to construct $U_{1}$ via adiabatic {method}. The Rabi frequency and detuning are selected as~\cite{t019}
\begin{equation}\label{eq7}
\Omega_{r1}=\left\{
\begin{aligned}
  & \Omega_{0}\Big[1-\cos\Big(\frac{\pi t}{\tau}\Big)\Big],& 0\leq t<\tau \\
 &\Omega_{0}\Big[1+\cos\Big(\frac{\pi (t-\tau)}{\tau}\Big)\Big]&\tau\leq t<2\tau\\
\end{aligned}
\right.
\end{equation}
and
\begin{equation}\label{eq8}
\Delta(t)=\left\{
\begin{aligned}
  & \Delta_{0}\Big[1+\cos\Big(\frac{\pi t}{\tau}\Big)\Big],& 0\leq t<\tau \\
 &\Delta_{0}\Big[\cos\Big(\frac{\pi (t-\tau)}{\tau}\Big)-1\Big]&\tau\leq t<2\tau\\
\end{aligned}
\right.
\end{equation}
to realize the population inverse $|0\rangle_{c}\rightarrow|r\rangle_{c}$ or $|1\rangle_{t}\rightarrow|r\rangle_{t}$ after time duration $2\tau$ (from time $0$ to $2\tau$). Then with parameters
\begin{equation}\label{eq9}
\Omega_{r1}=\left\{
\begin{aligned}
  & \Omega_{0}\Big[1-\cos\Big(\frac{\pi (t-2\tau)}{\tau}\Big)\Big],& 2\tau\leq t<3\tau \\
 &\Omega_{0}\Big[1+\cos\Big(\frac{\pi (t-3\tau)}{\tau}\Big)\Big]&3\tau\leq t<4\tau\\
\end{aligned}
\right.
\end{equation}
and
\begin{equation}\label{eq10}
\Delta(t)=\left\{
\begin{aligned}
  & \Delta_{0}\Big[1+\cos\Big(\frac{\pi (t-2\tau)}{\tau}\Big)\Big],& 2\tau\leq t<3\tau \\
 &\Delta_{0}\Big[\cos\Big(\frac{\pi (t-3\tau)}{\tau}\Big)-1\Big],&3\tau\leq t<4\tau\\
\end{aligned}
\right.
\end{equation}
$|r\rangle_{c}\rightarrow-|0\rangle_{c}$ or $|r\rangle_{t}\rightarrow-|1\rangle_{t}$ would be realized after time duration $2\tau$ (from time $2\tau$ to $4\tau$).
For control atom, using the Rabi frequency in Eq.~(\ref{eq7}) and detuning in Eq.~(\ref{eq8}) as step (i), and Rabi frequency in Eq.~(\ref{eq9}) and detuning in Eq.~(\ref{eq10}) as step (iii), $|0\rangle_{c}\rightarrow|r\rangle_{c}\rightarrow|0\rangle_{c}$ is realized within time duration $4\tau$. Here laser phase is set as $\varphi=\pi/2$ in step (i) and $\varphi=-\pi/2$ in step (iii). For target atom, adopting the above Rabi frequencies and detunings used in steps (i) and (iii) successively as the whole step (ii) {and making} $\varphi=0$, after the time duration $4\tau$, $|1\rangle_{t}\rightarrow e^{i\pi}|1\rangle_{t}$ is realized. Note that the above evolution is valid when the adiabatic condition is met, {and the adiabatic phases are offset during the cyclic evolution between two eigenstates of $H(t)$ in Eq}.~(\ref{eq1}). Obviously, the total operation time is $8\tau$ in constructing adiabatic $U_{1}$.

\emph{Non-adiabatic case}. Rabi frequency is selected as one truncated Gaussian pulse described by $\Omega(t)=\Omega_{n}e^{-t^2/\sigma}$~\cite{t019,t020} to realize non-adiaboatic $U_{1}$ ({subscript $n$ denotes non-adiabatic case}). The corresponding Hamiltonian for control~(target) atom is $H_{c,(t)}=\Omega(t)/2(e^{i\phi}|r\rangle\langle0_{c},(1_{t})|+\text{H.c.})$. It also needs three steps to construct non-adiabatic $U_{1}$. One $\pi$ pulse is required in steps (i) and (iii), while one $2\pi$ pulse with laser phase $\varphi=\pi$ is required in step (ii)~\cite{t001}.

\begin{figure}
  \centering
  \includegraphics[scale=0.4]{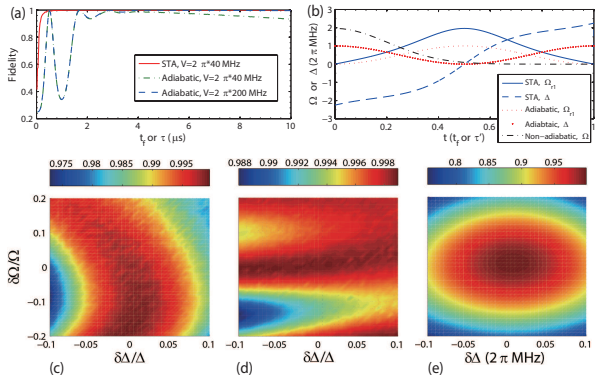}\\
  \caption{(a) Fidelities versus the single step time $t_{f}$ or $\tau$ in constructing $U_{1}$ of STA and adiabatic schemes. (b) {Rabi frequencies and detunings used for STA, adiabatic and non-adiabatic schemes within single step time [for step (i)]. Evolution time $t$ is in units of $t_{f}=1~\mu$s (STA and non-adiabatic cases) or $\tau'=2\tau=8~\mu$s (adiabatic case)}. Fidelities versus the relative deviations of Rabi frequency $\delta\Omega/\Omega$, detuning $\delta\Delta/\Delta$, or the deviation of detuning from the ideal case $\delta\Delta$ (in units of $2\pi$ MHz): (c) LR invariant scheme $U_{1}$ with operation time $T_{s}=4~\mu$s, RRI strength $V=2\pi\times40$ MHz; (d) Adiabatic scheme $U_{1}$ with operation time $T_{a}=32~\mu$s, RRI strength $V=2\pi\times200$ MHz; (e) Non-adiabatic scheme $U_{1}$ with operation time $T_{n}=4~\mu$s, RRI strength $V=2\pi\times40$ MHz.}\label{fig2}
\end{figure}

For STA $U_{1}$, setting total operation time as $T_{s}=4t_{f}=4~\mu$s ({subscript $s$ denotes STA case}), from Eq.~(\ref{eq3}), the corresponding maximum Rabi frequency and detuning are $\Omega_{s}\approx2\pi\times1.96$ MHz and $\Delta_{s}\approx2\pi\times2.25$ MHz, respectively. To make a comparison among the STA, adiabatic and non-adiabtaic $U_{1}$, it is rational to set the same maximum Rabi frequency and same RRI $V$ in non-adiabatic and adiabatic cases with STA scheme. For $\Omega_{n}=\Omega_{s}$ and $V=2\pi\times 40$ MHz, to construct non-adiaboatic $U_{1}$, it needs $\sigma\approx0.08293$ in steps (i) and (iii) to achieve the $\pi$ pulse, and $\sigma\approx0.33171$ in step (ii) to achieve the $2\pi$ pulse ($\sigma$ in units of time square). Total operation time $T_{n}$ in non-adiabatic scheme also equals $T_{s}$. However, in adiabatic scheme, the maximum Rabi frequency $\Omega_{a}$ ({subscript $a$ denotes adiabatic case}) is not suitable to be $2\pi\times1.96$ MHz, because a very strong RRI strength $V$ should be provided to ensure the adiabatic evolution quality. {As shown in Fig}.~\ref{fig2}(a), for a smaller maximum Rabi frequency $\Omega_{a}=2\Omega_{0}=2\pi$ MHz, detuning $2\Delta_{0}=2\pi$ MHz, and RRI strength $V=2\pi\times 40$ MHz, the fidelity of adiabatic $U_{1}$ decreases with increasing $\tau$ while it performs qualifiedly when $V=2\pi\times 200$ MHz. Thus for $\Omega_{a}>2\pi$ MHz, {it needs a larger $V>2\pi\times 200$ MHz to ensure the Rydberg blockade}. In Fig.~\ref{fig2}(a), the operation time needs to be $T_{a}=8\tau=32~\mu$s to construct well performing adiabatic $U_{1}$ with $V=2\pi\times200$ MHz, while it is enough for STA $U_{1}$ with operation time $4~\mu$s and $V=2\pi\times40$ MHz. Generally speaking a longer evolution time should result in a more stable and higher fidelity in adiabatic evolution, but it is unconventional for $V=2\pi\times40$ MHz in adiabatic scheme. Here we consider, for increasing time duration, the fidelity of adiabatic evolution decreases visibly due to the accumulated non-Rydberg blockade effect. And to modify this phenomenon, a larger RRI is needed. Fidelity in the letter {is defined as} $F=|\langle\psi_{ideal}|\psi(t)\rangle|^{2}$.

{Taking step (i) as an example, we show the Rabi frequencies and detunings (in units of $2\pi$ MHz) used for STA, adiabatic and non-adiabatic schemes within single step time in Fig.~\ref{fig2}(b), where the evolution time $t$ is in units of $t_{f}=1~\mu$s (for STA and non-adiabatic cases) or $\tau'=2\tau=8~\mu$s (for adiabatic case)}.

In Figs.~\ref{fig2}(c), \ref{fig2}(d) and \ref{fig2}(e), {we show the simulated fidelities of the three kinds of $U_{1}$ versus the corresponding relative Rabi frequency deviation (imprecision error) $\delta\Omega_{a,s(n)}/\Omega_{a,s(n)}$ and detuning deviation $\delta\Delta/\Delta$, where $\delta\Omega_{a,s(n)}=[\Omega_{a,s(n)}'-\Omega_{a,s(n)}]$ and $\Omega_{a,s(n)}'$ is the practical Rabi frequency considering fluctuation, here the same calculation to $\delta\Delta$}. Note $\delta\Delta$ in {Fig}.~\ref{fig2}(e) is the detuning deviation from ideal detuning (zero for resonance) in units of $2\pi$ MHz.  Obviously, the STA $U_{1}$ is more robust against the control parameter variations than non-adiabatic $U_{1}$ as shown in Figs.~\ref{fig2}(c) and \ref{fig2}(e). Besides, the STA $U_{1}$ is also comparable to adiabatic $U_{1}$ in the robustness and needs shorter operation time and smaller vdW interaction strength, which {are} shown in Figs.~\ref{fig2}(c) and \ref{fig2}(d).

Note that our scheme is valid and feasible under two fulfilled conditions. One is the efficient Rydberg blockade condition $V\gg\Omega_{r1}$; The other is about tiny decay, i.e., $t_{f}~(\tau)\ll\tau'$, where $\tau'$ denotes the lifetime of Rydberg state. In alkali atoms~\cite{t021,t022,t023}, it is shown that the lifetime of a Rydberg state with principle quantum number $n>70$ can be larger than $100~\mu$s and the RRI strength between two atoms are above $2\pi\times100$ MHz, when the atoms are separated less than $5~\mu$m. Thus the RRI strength $V=2\pi\times40$ MHz in our scheme is available and also the operation time $4~\mu$s $\ll100\mu$s. To show the influence of decay, we now analyse the fidelity of $U_{1}$ in consideration of spontaneous emission with the Lindblad master equation
\begin{equation}
  \dot{\rho}(t)=i[\rho(t),H(t)]+\sum_{k=1}^{2}\Big[L_{k}\rho L_{k}^{\dag}-\frac{1}{2}(L_{k}^{\dag}L_{k}\rho+\rho L_{k}^{\dag}L_{k})\Big],
\end{equation}
where $H(t)$ is the total Hamiltonian, $\rho(t)$ is the density operator of system state. $L_{k}~(k=1,2)$ are the Lindblad operators as $L_{1} = \sqrt{\gamma_{r0}}|0\rangle_{c}\langle r|,~L_{2} =\sqrt{\gamma_{r1}}|1\rangle_{t}\langle r|$, where $\gamma_{r0}$ and $\gamma_{r1}$ are the spontaneous emission rates of control and target atoms, respectively. For brief discussion, we assume $\gamma_{r0}=\gamma_{r1}=\gamma$. Considering the lifetime of Rydberg state is normally larger than $100~\mu$s, it is rational to set $\gamma=0.01$ MHz. Then the fidelities of $U_{1}$ in STA, adiabatic, and non-adiabatic schemes are $98.17\%,~87.70\%,~98.21\%$, respectively. Therefore, the STA scheme still performs well against decay. Note the condition $\Omega_{r1}\gg V_{1}$ cannot be satisfied at the begining and ending of the evolution process for constructing multi-qubit CPGs, because $\Omega_{r1}$ begins and ends up with zero. However, it does not have much influence on the STA scheme. For example, the fidelity of constructing three-qubit $\pi$ CPG is still 95.98\% for $V_{1}=2\pi\times0.1$ MHz. Here small $V_{1}$ could be achieved with longer separated distance between target atoms than the distance between control and target atoms. Condition $\Omega_{r1}\gg V_{1}$ can also be met {by} adopting larger $V$ of {RRI} between control and target atoms, {which allows the} larger $\Omega_{r1}$~\cite{t0181}.

In conclusion, {one} scheme to construct multi-qubit arbitrary-phase CPGs is presented based on {STA of} LR invariant theory. We take two-qubit $\pi$ CPG as an example for numerical simulation. {The results show that our scheme is more robust against control parameter variations than non-adiabatic case, and evolves faster than adiabatic case which also leads to a good robustness against decay}. Furthermore, comparing with the adiabatic case, smaller RRI strength is enough for the STA scheme.

This work was supported by National Natural Science Foundation
of China under Grant No. 11804308.


\begin{thebibliography}{999}
\bibitem{t001} D. Jaksch, J. I. Cirac, P. Zoller, S. L. Rolston, R. C\^{o}t\'{e}, and M. D. Lukin. Phys. Rev. Lett. {85}, 2208 (2000).
\bibitem{t002} M. Saffman, T. G.Walker, and K. M{\o}lmer. Rev. Mod. Phys. {82}, 2313 (2010).
\bibitem{t0021} S.-L. Su, E. Liang, S. Zhang, J. J. Wen, L. L. Sun, Z. Jin, and A. D. Zhu, Phys. Rev. A {93}, 012306 (2016).
\bibitem{t003} E. Urban, T. A. Johnson, T. Henage, L. Isenhower, D. D. Yavuz, T. G. Walker, and M. Saffman. Nat. Phys. {5}, 110 (2009).
\bibitem{t004} A. Ga\"{e}tan, Y. Miroshnychenko, T. Wilk, A. Chotia, M. Viteau, D. Comparat, P. Pillet, A. Browaeys, and P. Grangier. Nat. Phys. {5}, 115 (2009).
\bibitem{t005} L. Isenhower, E. Urban, X. L. Zhang, A. T. Gill, T. Henage, T. A. Johnson, T. G. Walker, and M. Saffman. Phys. Rev. Lett. {104}, 010503 (2010).
\bibitem{t006} K. Bergmann, H. Theuer, and B. W. Shore. Rev. Mod. Phys. {70}, 1003 (1998).
\bibitem{t007} D. M{\o}ller, L. B. Madsen, and K. M{\o}lmer. Phys. Rev. Lett. {100}, 170504 (2008).
\bibitem{t0071} D. D. B. Rao, and K. M{\o}lmer. Phys. Rev. A {89}, 030301(R) (2014).
\bibitem{t009} M. V. Berry. J. Phys. A {42}, 365303 (2009).
\bibitem{t010} X. Chen, I. Lizuain, A. Ruschhaupt, D. Gu\'{e}ry-Odelin, and J. G. Muga. Phys. Rev. Lett. {105}, 123003 (2010).
\bibitem{t011} H. R. Lewis, and W. B. Riesenfeld. J. Math. Phys. {10}, 1458 (1969).
\bibitem{t0110} X. Chen, A. Ruschhaupt, S. Schmidt, A. del Campo, D. Gu\'{e}ry-Odelin, and J. G. Muga. Phys. Rev. Lett. {104}, 063002 (2010).
\bibitem{t0120} A. Ruschhaupt, X. Chen, D. Alonso, and J. G. Muga. New J. Phys. {14}, 093040 (2012).
\bibitem{t012} X. Chen, E. Torrontegui, and J. G. Muga. Phys. Rev. A {83}, 062116 (2011).
\bibitem{t013} T. Abad, and K. M{\o}lmer. Phys. Rev. A {98}, 022324 (2018).
\bibitem{t016} C. P. Yang, Q. P. Su, F. Y. Zhang, and S. B. Zheng. Opt. Lett. {39}, 3312 (2014).
\bibitem{t017} T. Beth, and M. R\"{o}tteler. Quantum Information (Springer, Berlin, 2001), Vol. 173, Ch. 4, p. 96.

\bibitem{t014} M. V. Berry. Proc. R. Soc. Lond. A {392}, 45 (1984).
\bibitem{t015} S.-L. Su, Y. Gao, E. Liang, and S. Zhang. Phys. Rev. A {95}, 022319 (2017).
\bibitem{t0181} M. Saffman, and K. M{\o}lmer. Phys. Rev. Lett. {102}, 240502 (2009).
\bibitem{t019} Z. T. Liang, X. Yue, Q. Lv, Y. X. Du, W. Huang, H. Yan, and S. L. Zhu. Phys. Rev. A {93}, 040305 (2016).
\bibitem{t020} C. Zu, W.-B. Wang, L. He, W.-G. Zhang, C.-Y. Dai, F. Wang, and L.-M. Duan. Nature (London) {514}, 72 (2014).
\bibitem{t021} M. Saffman, and T. G. Walker. Phys. Rev. A {72}, 022347 (2005).
\bibitem{t022} A. Browaeys, D. Barredo, and T. Lahaye. J. Phys. B {49}, 152001 (2016).
\bibitem{t023} P. Z. Zhao, X. D. Cui, G. F. Xu, E. Sj\"{o}qvist, and D. M. Tong. Phys. Rev. A {96}, 052316 (2017).
\end{thebibliography}
\end{document}